\documentclass[twocolumn,amsmath,floatfix,prb,aps]{revtex4-1}

\usepackage{xcolor}
\usepackage{amsmath}
\usepackage{pifont}   % Ding symbols
\usepackage{graphicx} % Include figure files
\usepackage{dcolumn}  % Align table columns on decimal point
\usepackage{bm}       % bold math
\usepackage{amsfonts} % Some more fonts
\usepackage{amssymb}  % More symbols
\usepackage{multirow} % Table functions
\usepackage{romannum}
\usepackage{siunitx}
\usepackage{mathtools}

\DeclarePairedDelimiter\floor{\lfloor}{\rfloor}

    \usepackage{braket}
    \usepackage{nicefrac}
    \usepackage{bm}
    \usepackage{physics}

    \renewcommand{\v}[1]{\bm{\mathrm{#1}}}
    \newcommand{\m}[1]{\bm{\mathsf{#1}}}

\usepackage[english]{babel}
\usepackage[autostyle, english = american]{csquotes}
\MakeOuterQuote{"}

\begin{document}

%%%% User-defined commands %%%%
\newcommand{\ba}{{\bf a}}
\newcommand{\BB}{{\bf b}}
\newcommand{\bd}{{\bf d}}
\newcommand{\br}{{\bf r}}
\newcommand{\bp}{{\bf p}}
\newcommand{\bk}{{\bf k}}
\newcommand{\bg}{{\bf g}}
\newcommand{\bj}{{\bf j}}
\newcommand{\bt}{{\bf t}}
\newcommand{\bv}{{\bf v}}
\newcommand{\bu}{{\bf u}}
\newcommand{\bq}{{\bf q}}
\newcommand{\bG}{{\bf G}}
\newcommand{\bP}{{\bf P}}
\newcommand{\bJ}{{\bf J}}
\newcommand{\bK}{{\bf K}}
\newcommand{\bL}{{\bf L}}
\newcommand{\bR}{{\bf R}}
\newcommand{\bS}{{\bf S}}
\newcommand{\bT}{{\bf T}}
\newcommand{\bQ}{{\bf Q}}
\newcommand{\bA}{{\bf A}}
\newcommand{\bH}{{\bf H}}

\newcommand{\bdel}{\boldsymbol{\delta}}
\newcommand{\bsig}{\boldsymbol{\sigma}}
\newcommand{\beps}{\boldsymbol{\epsilon}}
\newcommand{\bnu}{\boldsymbol{\nu}}
\newcommand{\bnab}{\boldsymbol{\nabla}}
\newcommand{\bGam}{\boldsymbol{\Gamma}}

\newcommand{\bgt}{\tilde{\bf g}}
\newcommand{\brh}{\hat{\bf r}}
\newcommand{\bph}{\hat{\bf p}}

\title{Volkov-Pankratov states in a 2d material: excited states of a structural soliton}

\author{S. Theil$^1$}
\author{R. Gupta$^{2}$}
\author{F. Wullschl\"ager$^3$}
\author{B. Meyer$^3$}
\author{S. Sharma$^4$}
\author{S. Shallcross$^4$}
\email{shallcross@mbi-berlin.de}
\affiliation{1 Lehrstuhl f\"ur Theoretische Festk\"orperphysik, Staudtstr. 7-B2, 91058 Erlangen, Germany}
\affiliation{2 School of Physics and CRANN, Trinity College, 2, Dublin, Ireland}
\affiliation{3 Interdisciplinary Center for Molecular Materials (ICMM) and Computer Chemistry Center (CCC), Friedrich-Alexander-Universit\"at Erlangen-N\"urnberg (FAU), N\"agelsbachstra{\ss}e~25, 91052 Erlangen, Germany}
\affiliation{4 Max-Born-Institute for Non-Linear optics, Max-Born Strasse 2A, 12489 Berlin, Germany}

\date{\today}

\begin{abstract}
We show that partial dislocations, defects that naturally arise in bilayer graphene, host an analogue of the mass inversion Volkov-Pankratov (VP) states, the spectrum of excited states at a topologically non-trivial interface. In contrast to the dislocation states arising from the change in valley Chern index, such states (i) exist both with and without an interlayer bias, i.e. for metallic as well as insulating bulk and (ii) have distinct electron and hole bound states, whose number is proportional to the dislocation width. Recently discovered at heterojunctions of 3d topological insulators, their existence at the partial dislocations of a 2d material opens a rich structural playground for their investigation, and we demonstrate that the dislocation type, i.e. screw or edge, as well as the dislocation width, both play a decisive role in the creation of a topological spectrum of exited states.
\end{abstract}

\maketitle

%\section{Introduction}

{\it Introduction}: Bilayer graphene hosts perhaps the most remarkable physics of extended defects of any material in nature\cite{Bistritzer2011a, dai_twisted_2016,jain_structure_2016,Cao2018a,Cao2018b,ram18,Xu2019,san-jose_helical_2013}, including both two dimensional twist defects and one dimensional partial dislocations\cite{Alden2013,Butz2014,kiss15,shall17,2,3,8,12,4,6,9}. These latter defects consist of lines across which the stacking type of the bilayer changes from AB to BA, and can arise both as wandering and intersecting dislocation lines, as for example in the dense partial dislocation network found in bilayer graphene grown epitaxially on SiC ($000\overline{1}$)\cite{Alden2013,Butz2014}, or in a highly ordered form in minimally twisted bilayer graphene, which relaxes to a $C_3$ symmetric network of pure screw partial dislocations\cite{dai_twisted_2016,jain_structure_2016}.

At these line defects is revealed the non-trivial topological character of bilayer graphene: AB and BA stacked bilayers have differing valley Chern numbers and, in an applied interlayer bias, this generates a pair of topologically protected gap-crossing boundary states on the partial dislocations connecting regions of AB and BA stacking\cite{zhang_valley_2013,Huang2018,rickhaus_transport_2018}. Such states are important as they can both decisively influence transport properties, to the extent of changing an insulating material to a metallic one\cite{shall17}, as well as leading to a remarkable bias controllable Fermiology\cite{fleischmann_perfect_2020} and a corresponding rich magneto-transport physics\cite{rickhaus_transport_2018,de20} in minimally twisted bilayer graphene.

Volkov-Pankratov (VP) states, first predicted in a series of pioneering papers by Volkov and Pankratov more than 25 years ago, are the spectrum excited states found at a gap inversion semiconductor interface\cite{vol85,pan87}, and represent a rare opportunity to create designed topological states via interface manipulation. The VP spectrum has recently been predicted and observed in heterojunctions of topological insulators\cite{P1,P2,P3,P4,P5,P6}, and subsequently predicted to exist at the edge of graphene nanoribbons. However, despite the wide occurrence of topological insulators in condensed matter, the VP spectrum remains largely unexplored.

Here we show that partial dislocations in bilayer graphene host a two dimensional analogue of the Volkov-Pankratov (VP) states. In contrast to the bound states arising from the changing valley Chern index across a partial, which exist independent of the structural details of the transition from AB and BA stacking, these states are revealed only for sufficient width $l_c$ of the partial dislocation, and increase in number roughly as $\sim\floor{l/l_\perp}$ with $l_\perp = \hbar v_F/t_\perp$ the intrinsic length scale of bilayer graphene ($t_\perp$ is the interlayer hopping strength). For sufficient $l$ a rich physics of VP states is found, influenced both by the edge versus screw character of the partial dislocation, as well as the presence or absence of an interlayer bias.

For pure screw partial dislocations, of the type found in minimally twisted bilayer graphene, the critical $l$ for observing VP states is $l_c\sim 4$~nm, that our structural optimisation calculations suggest lie below the partial dislocation widths of this material. However, partial dislocations of mixed edge and screw character are predicted to host VP states for widths exceeding $l_c\sim1$~nm, well below typical dislocation widths. Such states should therefore be able to be observed in partial dislocation networks found in bilayer graphene grown epitaxially on SiC ($000\overline{1}$).

%%%%%%%%%%%%%%%%%% MODEL

{\it Hamiltonian}: Our method will consist of (i) treating bilayer graphene in the Dirac-Weyl approximation and (ii) a geometry of straight partial dislocations parallel to the armchair ($y$) direction. The corresponding Hamiltonian is 

\begin{equation}
H(x)=\begin{pmatrix}
(\sigma_x p_x + \sigma_y p_y) +\Delta & S(x)\\
S^\dagger(x) & (\sigma_x p_x - \sigma_y p_y) - \Delta
\end{pmatrix}
\label{H}
\end{equation}
where the spatially varying interlayer interaction is denoted $S(x)$ and we have
employed a convenient set of units in which energy is measured in terms of the interlayer hopping parameter $t_\perp$ and length in terms of $l_\perp = \hbar v_F/t_\perp$. As a model of the partial dislocation we introduce a function $\Delta \bu(x) = \bu_2(x)-\bu_1(x)$ that locally shifts the two layers of the initially AB stacked bilayer ($\bu_{1,2}$ represent the in-plane deformation field applied to each layer), and describe a partial dislocation by the form

\begin{equation}
\Delta \bu(x) = \frac{1}{2}\BB_i \left(1+\tanh(\frac{L(x-x_0)}{w})\right)
\label{X}
\end{equation}
where $w$ the partial dislocation width, $x_0$ the position of the dislocation core, $\BB_i$ one of the three possible partial Burgers vectors of graphene, $\BB_1 = (1/2,1/(2\sqrt{3}))a$, $\BB_2 = (0,-1/\sqrt{3})a$, and $\BB_3 = (-1/2,1/(2\sqrt{3}))a$. For the Burgers vector $\BB_2$ the corresponding shift function for a partial dislocation is shown in Fig.~\ref{fig1}. To obtain the interlayer potential $S(x)$ of the bilayer Hamiltonian Eq.~\eqref{H} from the interlayer deformation field Eq.~\eqref{X} we employ a standard $\pi$-orbital tight-binding Hamiltonian along with a continuum approach previously used for both the twist bilayer\cite{fleischmann_perfect_2020,Rost2019} and realistic partial dislocation networks\cite{Weckbecker2019}. In the SI we provide further details of this methodology. The interlayer stacking potential can, without loss of generality, be uniquely decomposed into stacking potentials of the three principle stacking types AA, AB, and BA, as

\begin{equation}
S(x) = \begin{pmatrix}
V_{AB}(x) & V_{AA}(x) \\
V_{AA}(x) & V_{AB}(x)
\end{pmatrix}
\end{equation}

As a partial dislocation represents a change in the stacking order of the bilayer, it will prove convenient to rewrite these potentials as a "stacking difference" potential
$S_z = \frac{1}{2}(V_\mathrm{AB}-V_\mathrm{BA})$, an "average stacking" potential $S_0 = \frac{1}{2} (V_\mathrm{AB} + V_\mathrm{BA})$, with the AA component of stacking remaining unchanged as $S_x = V_\mathrm{AA}$. Employing Pauli matrices we can now write the interlayer potential as

\begin{equation}
S(x) = \sigma_0 S_0(x) + \sigma_x S_x(x) + \sigma_z S_z(x)
\label{S1}
\end{equation}

\begin{figure}
\centering
\includegraphics[width=0.45\textwidth]{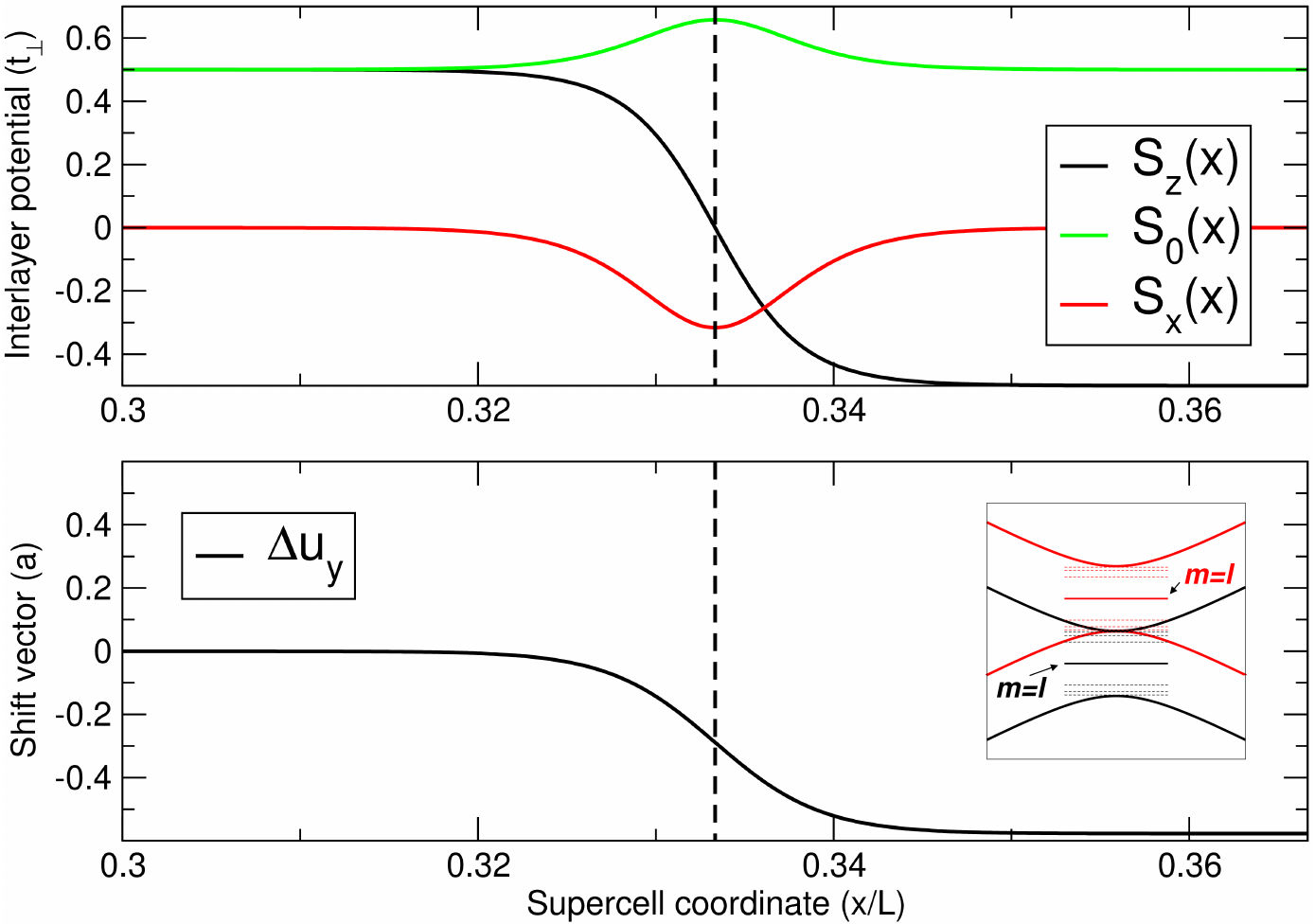}
\caption{Interlayer potentials (top panel) and interlayer shift function (bottom panel) for a pure screw partial dislocation (a Burgers vector of $(0,-\frac{1}{\sqrt{3}})a$ with dislocation line along the armchair $\hat{\v y}$ direction). The interlayer potentials are expressed in terms of a stacking difference potential ($S_z$) that changes sign from $+1/2$ to $-1/2$ across the dislocation, and a "stacking average" potential ($S_0$) that remains roughly constant across the dislocation. The deviation represents the appearance of AA stacking in the dislocation core, given by $S_x$. The inset panel illustrates the decomposition of the band structure of bilayer graphene into two gapped sets of bands along with midgap (labelled $m=l$) states and associated VP spectrum created by the sign change of $S_z$.}
\label{fig1}
\end{figure}

%%%%%%%%%%%%%%%% DERIVATION FOR SPECIAL CASE

{\it Pure screw partial dislocations}: For a partial dislocation aligned along the $y$ (i.e armchair) direction the Burgers vector $\BB_2$ represents a {\it pure screw} partial dislocation, i.e. a Burgers vector parallel to the dislocation line. The remaining two Burgers vectors represent mixed partial dislocations, with Burgers vectors neither parallel (screw) nor perpendicular (edge) to the dislocation line. The $\BB_2$ partial is particularly relevant as for hexagonal lattices the twist bilayer relaxes to a partial dislocation network formed of pure screw dislocations in the three armchair directions. For this geometry reflection symmetry about the dislocation line ensures that all interlayer potentials are real valued (a statement proved in the SI), and this allows for an analytical treatment of the emergence of Volkov-Pankratov states, which we now describe. Introducing a second set of $\tau$ Pauli matrices to describe the layer degree of freedom we may write the Hamiltonian Eq.~\eqref{H} as

\begin{equation}
H=\tau_0\otimes\sigma_x p_x + \tau_z\otimes\sigma_y p_y + \tau_z\otimes \sigma_0 \Delta + \tau_x \otimes S(x)
\label{HH}
\end{equation}
An $R_y(\pi/2)$ rotation in $\tau$-space followed by a gauge transformation \(U_2=\textrm{Diag}(\Phi_-,\Phi_+)\) with $\Phi_\pm (x) = e^{\pm\mathrm{i}\int_0^x S_x(x')\mathrm{d}x'}$ transforms Eq.~\eqref{HH} to

\begin{equation}
H_1=\begin{pmatrix}
\sigma_x p_x + \sigma_0 S_0+\sigma_z S_z & -(\sigma_y    p_y +\Delta) \Phi_-^2\\
-(\sigma_y p_y  + \Delta)\Phi_+^2 & \sigma_x p_x -       \sigma_0 S_0 -\sigma_z S_z
\end{pmatrix}
\label{A}
\end{equation}
and by setting $p_y=\Delta=0$, i.e. gapless graphene and zero momentum parallel to the dislocation line, the off-diagonal blocks vanish and Hamiltonian decouples into a pair of 1d Dirac Hamiltonians, which are gapped as both posses mass terms $\sigma_z S_z$. It is a striking feature that for a gapless bulk the underlying electronic structure can be described by a pair of gapped mass inversion Dirac-Weyl Hamiltonians (the mass term changes sign as $S_z$, the AB versus BA stacking difference, changes sign across the dislocation). The physical origin of this is that in decoupling the Hamiltonian we have broken the system into two auxiliary Hamiltonians describing the pairs of gapped bands formed from the Dirac point bands and anti-bonding and bonding bands respectively, illustrated in the inset of Fig.~\ref{fig1}. We can replace the "average stacking" potential by its bulk value of $1/2$, an excellent approximation as the average of AB and BA stacking is roughly constant across a dislocation (see Fig.~\ref{fig1}), giving after a final rotation $R_x(\pi/2)$ in $\sigma$-space the standard Jackiw-Rebbi (JR) form

\begin{equation}
H_1'=
\begin{pmatrix}
\sigma_x p_x + \sigma_y S_z + 1/2 & 0\\
0 & \sigma_x p_x - \sigma_y S_z -1/2
\end{pmatrix}
\label{H1prime}
\end{equation}
with solutions

\begin{equation}
\phi_\pm  = \ket{\pm}_\sigma\otimes\ket{\mp}_\tau e^{-\int_0^x S_z(x') \mathrm{d}x'}
\end{equation}
where $\sigma_z \ket{\pm}_\sigma = \pm \ket{\pm}_\sigma$ and $\tau_z \ket{\pm}_\tau = \pm \ket{\pm}_\tau$.

The presence of JR zero modes has been previously noted in an analysis of helical networks in twist bilayer graphene\cite{dim18}. However, by themselves these JR states are too far (200~meV) from the Dirac point to be of interest. As we now show, however, these represent the ground states of a series of excited states, the Volkov-Pankratov spectrum, whose energies can approach the Dirac point and are therefore of interest. To reveal this we consider a model stacking difference potential 

\begin{equation}
S_z(x)=\frac{1}{2} \tanh(\frac{x}{2l}),
\end{equation}
where \(l\) is the length scale of the partial dislocation which is chosen to be integer (i.e. in physical dimensions a multiple of $l_\perp$).
As shown in supplementary, for this potential form Eq.~\eqref{H1prime} reduces to two coupled P\"oschl-Teller Hamiltonians with a spectrum of states

\begin{equation}
E=\pm\frac{1}{2}\pm\frac{\sqrt{l^2-m^2}}{2l}
\label{E}
\end{equation}
where $l$ and $m$ are integers with $m \le l$. The eigenvalues with $l<m$ are the Volkov-Pankratov spectrum of the JR zero mode $l=m$ (indicated in the inset to Fig.~1). As we have measured $l$ in terms of $l_\perp=\hbar v_F/t_\perp$, for a partial dislocation with length scale \( l\), in dimensionful units, there are therefore \(\sim 2\lfloor l/l_\perp\rfloor +1\) Volkov-Pankratov states. As $l_\perp$ is of the order of $10\si{\angstrom}$ partial dislocations and the slower stacking transitions found in the twist bilayer would appear excellent candidates for observing VP states.

%%%%%%%%%%%%%%%%%%%% ATOMISTIC SIMULATION

\begin{figure}
\centering
\includegraphics[width=0.45\textwidth]{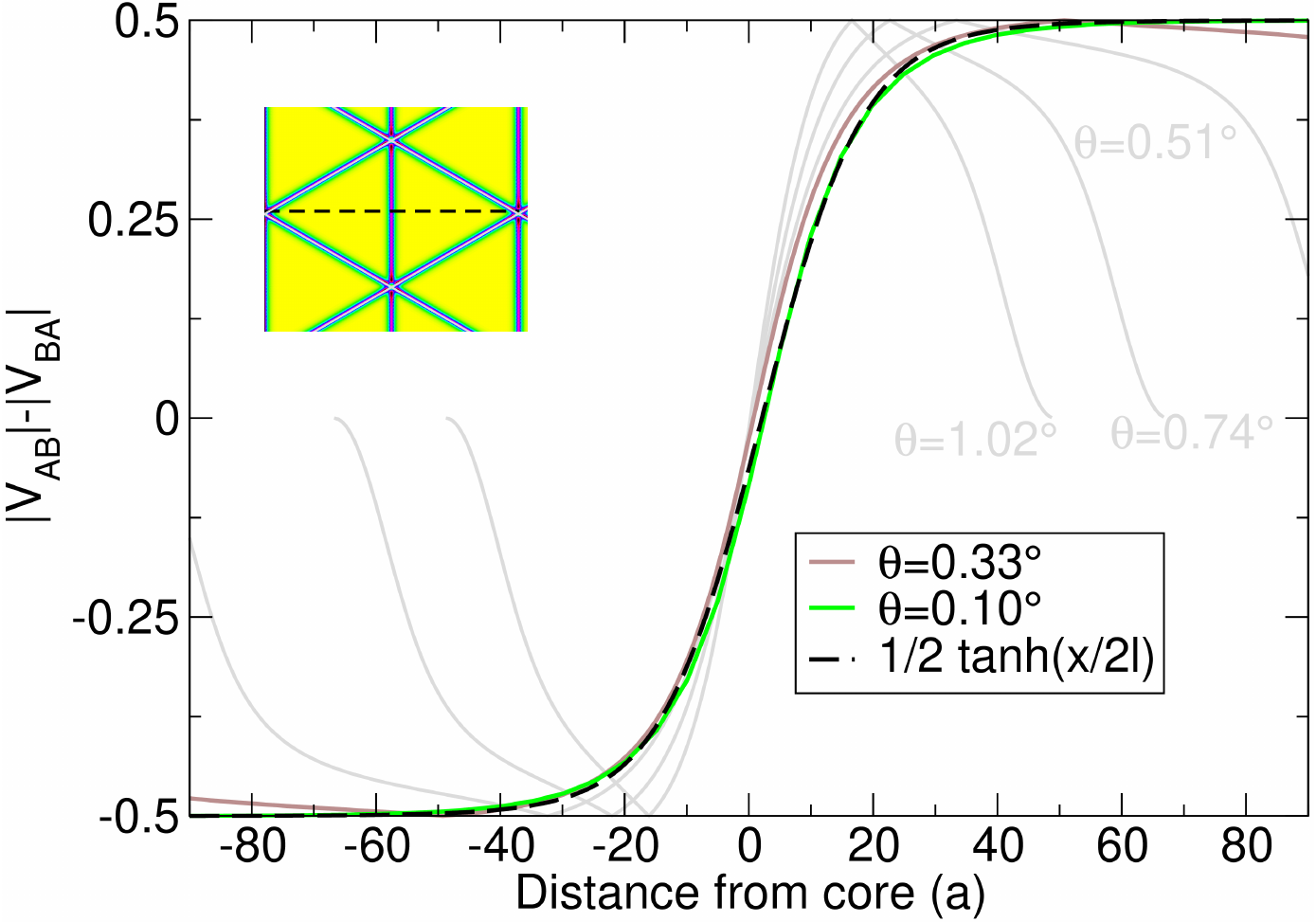}
\caption{Effective potential describing the difference of AB and BA stacking, derived from atomistic total energy calculations of the twist bilayer for a range of angles betwen $\theta=1.02^\circ$ and $\theta=0.10^\circ$. For $\theta < 0.33^\circ$ the minimally twisted bilayer is well described as a partial dislocation network, with further reduction in twist angle changing only the network scale. As can be seen, in this limit the dislocation potential function $\frac{1}{2}(V_{AB}-V_{BA})$ is very well described by the hyperbolic tangent function form given in Eq.~\eqref{X}. The inset displays the moir\'e unit cell along the line within this unit cell for which the potential is plotted.}
\label{fig2}
\end{figure}

{\it Dislocation structure of the minimally twisted bilayer}: Before embarking on numerical solution of Eq.~\ref{H} to investigate the existence of VP states without the restrictions necessary for an analytical derivation, we pause to consider the question of the form of the dislocation boundary. From Eq.~\eqref{E} it is apparent that the critical property for the existence of a VP spectrum is the width of the dislocation line. From experiment we know that for wandering partials created due to lattice mismatch in epitaxial growth on, for example, SiC (000$\overline{1}$), the typical partial dislocation width of $l=20a-40a$ (i.e. 5 to 10~nm)\cite{Alden2013,Butz2014}. For the pure screw relaxed twist bilayer, however, the detailed structure of the pure screw partial and the corresponding interlayer stacking potential has been less explored. We now address the question of the partial dislocation structure by performing atomistic calculations using the GAFF force field \cite{GAFF} for the C--C interactions within the graphene layers and the registry-dependent interlayer potential of Kolmogorov-Crespi \cite{KC2005}. From the atomic structure we extract, via bicubic interpolation, a form of the interlayer shift function (now $\br$-dependent) and from this in turn obtain the interlayer interaction. In Fig.~\ref{fig2} is shown the stacking difference potential $S_z$ revealing that (i) the stacking difference potential follows an almost perfect hyperbolic tangent form across the partial dislocation, and (ii) a corresponding dislocation width $l=21a$, similar to the dislocation widths found for wandering partials. Thus while the hyperbolic tangent form was deployed as a simplifying model in obtaining the VP spectrum, it turns out to almost perfectly describe the stacking difference field extracted from the realistic structure of pure screw partial dislocations in minimally twisted bilayer.

%%%%%%%%%%%%%%% VP STATES IN GENERAL

\begin{figure}
\centering
\includegraphics[width=0.49\textwidth]{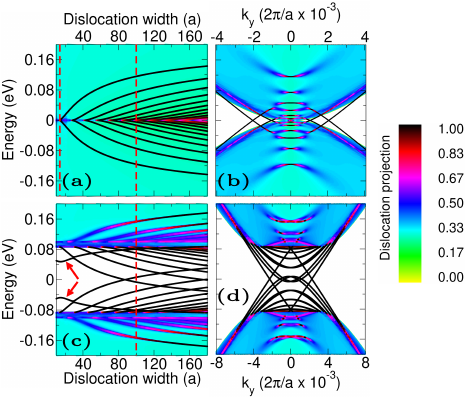}
\caption{Volkov-Pankratov (VP) states at a pure screw partial dislocation. In panel (a) is shown the dependence of the VP spectrum at $p_y=0$, plotted as a function of partial dislocation width $l$ (in units of the lattice parameter $a$) showing that as the dislocation width increases an increasing number of VP states are bound to the partial dislocation. The full band structure in the 1d Brillouin zone (partial with $l=100a$ indicated by the rightmost verical red line in panel (a)) reveals the VP states to exist only for small $p_y$ (b). Applying a finite interlayer bias opens a gap in the bilayer spectrum, with VP states now emerging symmetrically from the gap edge, as shown (c,d). In each panel the eigenvalue colour indicates the localization on the partial dislocations, defined as the region $\pm 0.08L$ either side of the dislocation defined at $1/3L$ and $2/3L$ in the supercell.}
\label{fig3}
\end{figure}

{\it Volkov-Pankratov states}: We now lift all the assumptions of our simplified derivation, for which Eq.~\eqref{H} must be solved numerically. Our setup will consist of two partial dislocations symmetrically placed at $x/L = 1/3$ and $x/L = 2/3$ in a unit cell of length $L=10000a$, with partial Burgers summing to zero to give a periodic system. 

\begin{figure}[t!]
\centering
\includegraphics[width=0.49\textwidth]{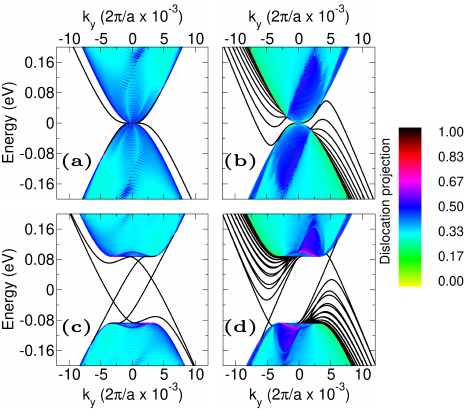}
\caption{Spectrum of Volkov-Pankratov states on a mixed edge/screw partial dislocation for widths of $l = 10a$, $20a$, $50a$, and $100a$ both for zero interlayer bias, panels (a-d), and a finite interlayer bias of 200\,meV, panels (e-h). Eigenvalue colour indicates the degree of localization on the partial dislocations as indicated by the colour bar.}
\label{fig4}
\end{figure}

We first probe the evolution with $l$ of the VP spectrum of pure screw partials (partial Burgers vector $\BB_1 = (0,-1/\sqrt{3})a$). In Fig.~\ref{fig3}a is displayed the VP spectrum at $p_y=0$ plotted versus partial width $l$. One observes a succession of states emerging symmetrically from zero energy as the partial dislocation width $l$ increases, beginning at $l_c=14a$ (indicated by the leftmost vertical red line). In agreement with the model calculation we find midgap Jackiw-Rebbi states at $t_\perp/2=200$~meV for all $l$ (not visible in the energy range of the plot). In panel (b) is displayed the band structure in the effective 1d Brillouin zone for a partial dislocation width of $100a$, with each point shaded according to its projection onto the partial dislocation. This reveals the VP states to exist only in a narrow range of momenta around $p_y=0$ (i.e., very close to the K point in the full two dimensional BZ), and to exhibit very little dispersion. Evidently, finite $p_y$ couples the electron and hole sectors destroying the VP states. 
Upon switching on an interlayer bias the VP spectrum now emerges symmetrically from the conduction and valence band edges (Fig.~\ref{fig2}c). The corresponding band structure (Fig.~\ref{fig2}d) shows that while the VP states emerge symmetrically from the electron and hole bulk spectrum they do not cross the gap, a feature also of the VP states found at topological heterojunctions and in the original work of Volkov and Pankratov on the SnTe/PbTe heterojunction. Note that the linear gap crossing states arising from the changing valley Chern index can also clearly be seen (indicated by the arrows in Fig.~\ref{fig2}c).

We now consider a mixed edge and screw partial dislocation, given by $\BB_1 = (1/2,1/\sqrt{3})a$. In contrast to the pure screw partial dislocation this geometry generates complex dislocation potentials, for which the analytical treatment given for the pure screw partial with real potentials evidently does not hold. It is thus an open question as to whether VP states exist in this more complex potential landscape of the mixed character partial dislocation.
As can be seen in panels (a-d) of Fig.~\ref{fig3} in dramatic contrast to the rather weak VP states found for the pure screw partial, this partial dislocation exhibits robust and dispersive skew symmetric ($p_y\to-p_y$ and $E\to-E$) VP states. The partial width for the onset of the VP spectrum is also significantly reduced; $5a$ as compared to $14a$ for the pure screw partial. 
No signature is seen of a JR state, or even a residual of this state, highlighting the dramatic change in bound state electronic structure that results from introducing edge character to the partial dislocation. 
Upon switching on an interlayer bias the VP spectrum splits with $p_y>0$ states attached to the valence band edge and $p_y <0$ states attached to the conduction band edge; again only the bands due to the changing valley Chern index cross the gap.

%%%%%%%% DISC.

{\it Discussion}: We have shown that partial dislocations in bilayer graphene host a rich physics of Volkov-Pankratov mass inversion states. Two remarkable differences exist in this 2d material manifestation of VP states as compared their previous observation in 3d materials. (i) While in 3d topological heterojunctions the presence of a gapped bulk is essential, in bilayer graphene VP states are more general, being found at stacking boundaries for both metallic as well as insulating bulk. (ii) The VP states are sensitive to the details of the partial dislocation structure, with dramatically different VP spectra found for pure screw  versus mixed edge/screw dislocations.
In minimally twisted bilayer graphene, which relaxes to an ordered network of pure screw partial dislocations, these states will constitute a second "Volkov-Pankratov network" of 1d states in addition to the "helical network" generated by the spatially changing Chern index. However, our calculations show that the critical dislocation width for the onset of VP states ($\sim 4$nm) is comparable to the dislocation width calculated for a relaxed $0.1^\circ$ twist bilayer, suggesting they represent a marginal case for this material. In contrast, we find that partial dislocations of mixed edge/screw character represent a case where the typical dislocation widths easily exceeds the critical width for the observation of VP states.
Taken together with their ease of study in a 2d material, which can be set in contrast to the less accessible VP spectrum that occurs at the interface of two 3d materials, then bilayer graphene would appear to represent perhaps the perfect material for the investigation and manipulation of Volkov-Pankratov states.

\section{Supplementary information}

\subsection{Continuum formalism}
\label{cont}

We employ for our continuum calculations the methodology described in Ref.~\onlinecite{Rost2019}. The idea of this approach is to consider a high symmetry system, lattice vectors $\v a_i$, basis vectors $\m \nu_\alpha$ and reciprocal lattice vectors $\v G_i$, to which a deformation is applied by changing the value of the tight-binding hopping function. The general two-centre tight-binding Hamiltonian for this situation then reads

\begin{equation}
H_{TB} = \sum_{\substack{\alpha  \beta \\ \br  \bdel}} t_{\alpha\beta}(\br,\bdel) c^\dagger_{\beta\br+\bdel} c_{\alpha\br}
\label{eqTB}
\end{equation}
where $\alpha$ and $\beta$ are combined atomic indices of the high symmetry system, i.e. layer index, basis atom index, angular momentum, and spin. The vector $\v r$ sums over all lattice sites having basis vector $\m\nu_\alpha$, with $\m \delta$ the hopping vector from $\v r$ to some distant site $\v r+\m \delta$ having basis vector $\m\nu_\beta$. The amplitude of the hopping matrix elements between these sites is given by $t_{\alpha\beta}(\br,\bdel)$ which for the purely high symmetry system (i.e., no deformation) would reduce to $t^{(0)}_{\alpha\beta}(\m\delta)$. The $\v r$ dependence thus encodes the spatial variation of hopping introduced by an applied deformation.

As shown in Ref.~\onlinecite{Rost2019} for any deformation field this Hamiltonian exactly maps onto the following continuum Hamiltonian

\begin{equation}
[H(\br,\v{p})]_{\alpha \beta} = \frac{1}{A_{\mathrm{UC}}} \sum_{\v{G}_j} [M_{j}]_{\alpha\beta} \:\eta_{\alpha\beta} (\br,\v{K}_j+ \v{p})
\label{simplified_hamiltonian}
\end{equation}
where the sum is over the reciprocal lattice vectors $\v G_j$ of the underlying high symmetry system, and $\bK_j = \bK_0 + \bG_j$ with $\bK_0$ some conveniently chosen reference momentum, which for bilayer graphene is one of the six K points. $A_{UC}$ is the area of the real space unit cell of the high symmetry structure. The function $\eta_{\alpha\beta}(\br,\bq)$ is the mixed space Fourier transform of the hopping function $t_{\alpha\beta}(\br,\bdel)$ defined in Eq.~\eqref{eqTB},

\begin{equation}
\eta_{\alpha\beta}(\br,\bq) = \int d\bdel\, e^{i\bq.\bdel} t_{\alpha\beta}(\br,\bdel)
\label{Ahop}
\end{equation}
Note the absence of any formal restriction on the spatial dependence of the change in hopping function matrix elements due to deformation. The deformation may thus both be perturbative, for example in the case of strain fields applied to each layer, as well as highly non-perturbative as is the case for interlayer deformations such as the introduction of twist faults or dislocations into a pristine AB stacked lattice as is considered here.
Finally, the "M matrices" are given by

\begin{equation}
[M_j]_{\alpha\beta} = e^{\mathrm{i}\v{G}_j\cdot(\bnu_\alpha-  \bnu_\beta)}
\end{equation}
and these, since they combine the real space basis vectors $\v \nu$ with the reciprocal lattice vectors $\v G_j$ of the high symmetry lattice this expression, contain the full information of the lattice and basis of the high symmetry system.
For further details we refer the reader to Ref.~\onlinecite{Rost2019} as well as to several recent applications of the method: to minimally twisted bilayer graphene\cite{fleischmann_perfect_2020}, partial dislocation networks\cite{kiss15,shall17,Weckbecker2019}, and in-plane deformation fields\cite{gupta19,gupta19a}.

For the case of interlayer deformations we consider a deformation field $\v u_i(\v r)$ applied to each layer $i$, with the difference $\Delta \v u (\v r) = \v u_2(\v r) - \v u_1(\v r)$ then describing the change in local stacking at each point $\v r$. The interlayer coupling blocks can also be obtained from Eq.~\eqref{simplified_hamiltonian} and, as shown by Rost {\it et al.}, the required Fourier transform Eq.~\eqref{Ahop} can be taken exactly to yield a general form for the interlayer interaction\cite{Rost2019} given by

\begin{equation}
[S(\v r,\v p)]_{\alpha\beta} = \frac{1}{A_{\mathrm{UC}}} 
\sum_j [M_j]_{\alpha\beta}\:e^{-\mathrm{i}\Delta \v{u}(\v{r})\cdot \v{G}_j}\:\hat{t}^{(0)}(\v{K}_j+\v{p})
\label{interlayer_coupling}
\end{equation}
where $\hat{t}^{(0)}(\v q) = \int d\bdel\, e^{i\bq.\bdel} t^{(0)}_{\alpha\beta}(\bdel)$ is the Fourier transform of the hopping envelope function of the high symmetry lattice.
The full Hamiltonian of the bilayer system is thus given by

\begin{equation}
H=\begin{pmatrix}
H^{(1)}(\v r, \v p) & S(\v r, \v p)\\
S^\dagger(\v r, \v p) & H^{(2)}(\v r, \v p)
\end{pmatrix}
\label{bilHam}
\end{equation}
For investigating dislocations in bilayer graphene we make a number of simplifying assumptions to this general expression. Firstly, we neglect the momentum dependence of the interlayer interaction, and thus this is given by

\begin{equation}
[S(\v r)]_{\alpha\beta} = \frac{1}{A_{\mathrm{UC}}} 
\sum_j [M_j]_{\alpha\beta}\:e^{-\mathrm{i}\Delta \v{u}(\v{r})\cdot \v{G}_j}\:\hat{t}^{(0)}(\v{K}_j)
\end{equation}
directly relating the local misregistry of the underlying high symmetry bilayer introduced by the deformation field $\Delta \v u(\v r)$ to the $2\times2$ interlayer interaction. As shown in Ref.~\onlinecite{Rost2019}, the momentum dependence of Eq.~\ref{interlayer_coupling}, while crucial for certain materials such as phosphorene, plays only a minor role in the interlayer interaction in graphene. We thus have to solve the Hamiltonian Secondly, for the diagonal blocks we (i) retain only linear order in $\v p$, i.e. we employ the Dirac-Weyl approximation, and (ii) we neglect the effect of in-plane strain fields on the two graphene layers. As shown in Ref.~\onlinecite{fleischmann_perfect_2020} in bilayer systems the change in the electronic structure is dominated by the interlayer deformation terms, and to a good approximation these terms can be ignored. 

The $C_3$ symmetry of graphene demands that each star of the translation group of momentum boosts encoded in the above equation is described by the same 3 "M matrices":

\begin{equation}
M_0=\begin{pmatrix}
1&1\\
1&1
\end{pmatrix},\qquad
M_\pm=\begin{pmatrix}
1&e^{\pm 2\pi\mathrm{i}/3}\\
e^{\mp 2\pi \mathrm{i}/3} & 1
\end{pmatrix}
\label{M}
\end{equation}
For this reason the interlayer interaction can be expressed as a sum of 3 distinct parts, with the most natural way to decompose the interlayer interaction in terms of the three stacking potentials associated with the high symmetry structures AB, BA and AA stacking, which have off-diagonal matrix structure $\sigma_+$, $\sigma_-$, and $\sigma_0$ respectively.

To evaluate the lattice sums in the expressions above we must specify the lattice geometry of the high symmetry system and we consider here a standard lattice geometry $\v a_1=(0,1)a$, $\v a_2 = (\frac{1}{2},\frac{\sqrt{3}}{2})a$ with basis vectors $\bnu_1=\v 0$, $\bnu_2=(1,1/\sqrt{3})a$, $\bnu_3=\v 0$, and $\bnu_4 = (1/2,1/(2\sqrt{3})$, where basis vectors 1 and 2 constitute the first layer basis and 3 and 4 the second layer basis. Using these vectors in Eq.~\eqref{simplified_hamiltonian} the diagonal blocks of the Hamiltonian are $v_F \sigma.\bp$ for the first layer and $v_F \sigma^\ast.\bp$ for the second layer.

In this work we consider straight, and not wandering, partial dislocations. The electronic structure problem is then effectively a 1d problem and we can write our final Hamiltonian as

\begin{equation}
H(x)=\begin{pmatrix}
(\sigma_x p_x + \sigma_y p_y) +\Delta & S(x)\\
S^\dagger(x) & (\sigma_x p_x - \sigma_y p_y) - \Delta
\end{pmatrix}
\label{H}
\end{equation}
where we have used the fact that we consider straight, not wandering, partial dislocation systems we which we align in the $y$ direction, and non-dimensionalized the Hamiltonian by measuring energies and lengths in terms of the constants \(t_\perp\) and \(l_\perp = \hbar v_\mathrm{F}/t_\perp\) respectively. The interlayer bias due to an applied layer perpendicular field $\pm\Delta$ has also been introduced. This expression is Eq.~1 of the main text.

\subsection{Tight-binding parameterization}

\begin{figure}[t!]
\centering
\includegraphics[width=0.45\textwidth]{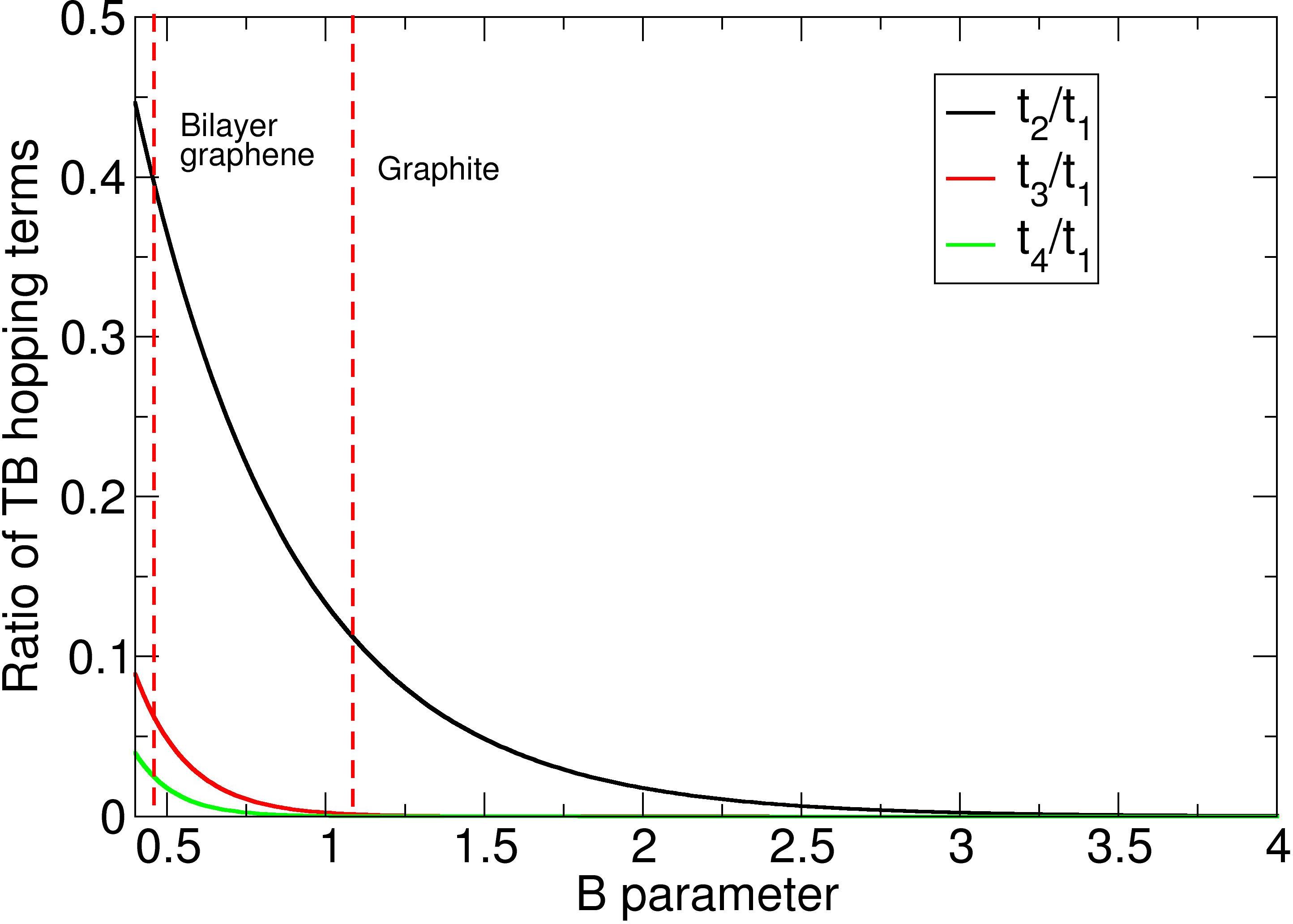}
\caption{The ratio of second, third and fourth nearest neighbour interlayer hopping to nearest neighbour hopping for values of \(B_\perp\) from 0 to 4. The second neighbour interlayer hopping remains relevant for \(B_\perp = 1.5-2\), while more distant hopping terms very quickly decay to zero.}
\label{hopping_range}
\end{figure}

For the tight-binding parameterization we will employ the H\"uckel (i.e. $\pi$-band only) approach and take the in-plane $t_\parallel^{(0)}$ and interlayer $t_\perp^{(0)}$ hopping functions to be parametrised by the Gaussian form

\begin{equation}
t^{(0)}({\m \delta}) = Ae^{-B {\m \delta}^2}.
\label{gauss}
\end{equation}
where \(A_\parallel\) and \(B_\parallel\) are chosen to give an in-plane nearest neighbour hopping of 2.8\(\:\)eV and the interlayer \(A_\perp\) and \(B_\perp\) are chosen such that the hopping between atoms directly coincident in the \(x-y\)-plane is 0.4\(\:\)eV.
By changing \(B_\perp\) we can interpolate between a "hard" fast decaying interlayer interaction (large \(B_\perp\)) and a "soft" slowly decaying one (small \(B_\perp\)). 
In all cases the \(A_\perp\) is modified so that the hopping between \(x-y\)-coincident atoms in each layer remains 0.4\(\:\)eV.
The effect this has on the first four nearest neighbour interlayer hopping terms in the AB bilayer is shown in Fig.~\ref{hopping_range}. 
For large values of \(B_\perp\) the interlayer interaction reduces to only nearest neighbour hopping, while at smaller value of $B_\perp$ the longer range hopping includes up to fourth nearest neighbour terms.
The ratio of next-nearest to nearest neighbour interlayer parameters in the Slonczewski-Weiss-McClure model employed in graphene and graphite \cite{mccann_electronic_2013,malard_probing_2007,dresselhaus_intercalation_2010} are indicated by the vertical lines. As can be seen, the parameter $B_\perp$ for the interlayer hopping must be $\sim 0.5$\si{\angstrom}$^{-2}$ to reproduce the nearest to next nearest neighbour hopping in the Slonczewski-Weiss-McClure model.

\begin{figure}
\centering
\includegraphics[width=0.45\textwidth]{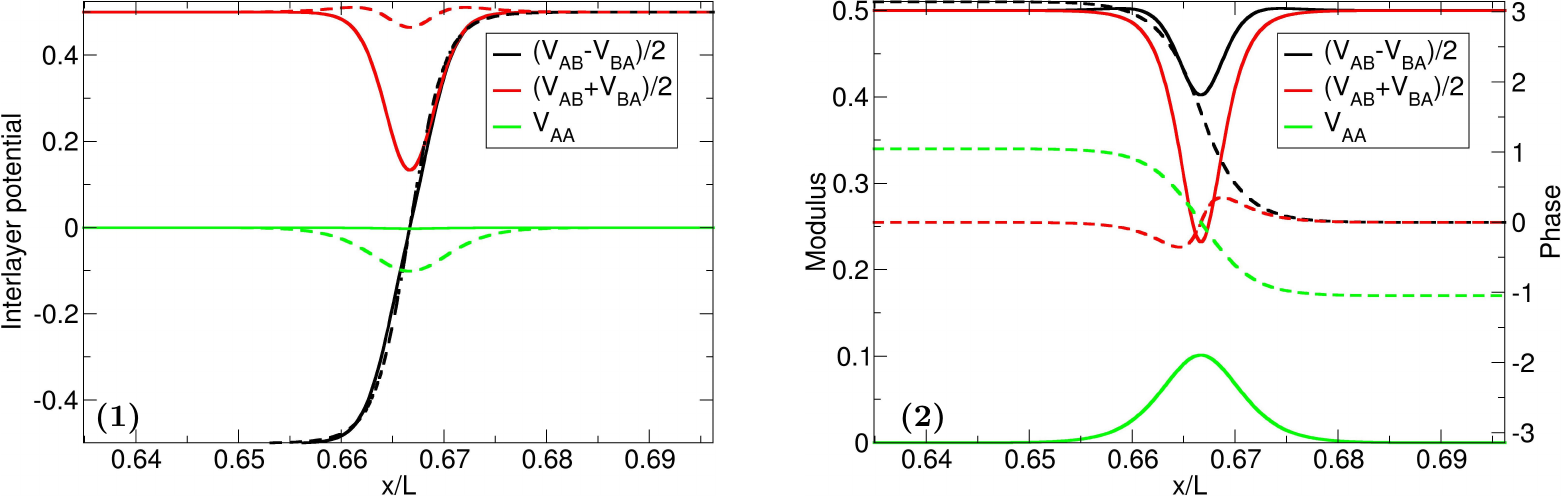}
\caption{Panel 1: Interlayer potentials for a pure shear partial dislocation. Dashed lines for the case of a slowly decaying "soft" interlayer interaction ($B_\perp=1.5$) and full lines for a fast decaying "hard" interaction ($B_\perp=4.0$). Note that in the latter case, the potential \(S_x = V_{AA}\) (see Eq.~\ref{S1}) is zero even at the dislocation centre, as the fast decay results in AA type stacking having no contribution to the interlayer interaction.
This fact also underpins the pronounced "depletion" of the average stacking potential (\(S_0 = (V_\mathrm{AB} + V_\mathrm{BA})/2\)) at the dislocation core.
In contrast, for the soft interlayer interaction there is appreciable \(V_\mathrm{AA}\) at the centre. 
Panel 2: Interlayer potentials for a partial dislocation of mixed edge and screw character ($B_\perp=1.5$). Here the potentials are complex valued with a phase change of \(\pi\) for the \(S_z\) stacking difference potential and \(2\pi/3\) for the \(S_x\)-potential}.
\label{pot2}
\end{figure}

\subsection{Method of numerical solution}

An effective approach to solving the continuum problem to use a basis formed from the single layer eigenstates\cite{fleischmann_perfect_2020}, which can be obtained by solving the single-layer blocks of the Hamiltonian as a pre-step to the full calculation

\begin{equation}
H_\mathrm{0}^{(n)}\ket{\Psi_{i\v k}^{(n)}} = \epsilon_{i \v k}^{(n)}\ket{\Psi_{i\v k}^{(n)}}
\end{equation}
where $n$ denotes the layer index.
We find a basis size of 3600 of the lowest energy states from each layer serve as a highly converged basis for solving the full bilayer problem. The matrix elements of $H$ in this basis are given by

\begin{equation}
[H]_{n'i'\v k ' ni\v k}=\delta_{n'i'\v k'ni\v k}\epsilon_{i \v k}^{(n)} + (1-\delta_{nn'})\mel{\Psi_{i'\v k'}^{(n')}}{S(x)}{\Psi_{i\v k}^{(n)}}
\end{equation}

\subsection{Atomic relaxation and extraction of stacking potentials}

Our calculational setup consists of the GAFF force field \cite{GAFF} for the C--C interactions within the graphene layers and the registry-dependent interlayer potential of Kolmogorov-Crespi \cite{KC2005}. For the ideal AB-stacked graphene bilayer we obtain  an equilibrium lattice constant of $a_0=2.441$\,{\AA} and an interlayer distance of $d_{\rm AB}=3.370$\,{\AA}. Shifting the graphene layers to AA stacking increases the layer separation to $d_{\rm AA}=3.597$\,{\AA} ($+0.227\si{\angstrom}$ as compared to AB stacking). The AA-stacked bilayer has a higher energy of 4.4\,meV per atom as compared to AB-stacking, corresponding to a stacking fault energy of $\gamma_{\rm AA} = 54.9$\,mJ/m$^2$. In SP stacking order the equilibrium distance of the graphene layers and the stacking fault energy are $d_{\rm SP}=3.390$\,{\AA} (+0.020\,\AA) and $\gamma_{\rm SP} = 7.1$\,mJ/m$^2$ (0.6\,meV per atom), respectively, in excellent agreement with ACFDT-RPA calculations of Srolovitz \emph{et al.} \cite{sor15}. Finally, a continuum vector field is then created from the atomic data by bicubic interpolation and this gives us directly the $\Delta \bu(\br)$ interlayer deformation field from which the interlayer interactions shown in Fig.~2 of the main text are obtained.

\section{Analytical solution for mass inversion bound states}
\label{model}

\subsection{Model}

We consider semi-infinite extended regions of AB and BA stacked bilayer graphene connected by a boundary along the armchair direction at \(x=0\). A continuous lattice implies that across this boundary the bilayer misregistry should change continuously by a partial Burgers vector. The Hamiltonian for this system is 

\begin{equation}
H=\begin{pmatrix}
v_\mathrm{F} \v{\sigma}\cdot\v{p} + \Delta& t_\perp S(x)\\
t_\perp S^\dagger(x) & v_\mathrm{F}\v{\sigma}^*\cdot\v{p}-\Delta
\end{pmatrix}=\begin{pmatrix}
v_\mathrm{F}(\sigma_x p_x + \sigma_y p_y) +\Delta& t_\perp S(x)\\
t_\perp S^\dagger(x) & v_\mathrm{F}(\sigma_x p_x - \sigma_y p_y) - \Delta
\end{pmatrix}.
\end{equation}
with the stacking potentials in the coupling blocks given by

\begin{equation}
S(x) = \begin{pmatrix}
V_\mathrm{AB}(x) & V_\mathrm{AA}(x)\\
V_\mathrm{AA}(x) & V_\mathrm{BA}(x)
\end{pmatrix}.
\label{interlayerNew}
\end{equation}

Note that we take the interlayer potentials to be momentum independent. By measuring energies and lengths in terms of the constants \(t_\perp\) and \(l_\perp = \hbar v_\mathrm{F}/t_\perp\) respectively we can non-dimensionalize this Hamiltonian.

We now rewrite \(S(x)\) in terms of three new potentials \(S_0\), \(S_x\) and \(S_z\)

\begin{eqnarray}
S(x) & = & \sigma_0 S_0+\sigma_x S_x+\sigma_zS_z \nonumber \\
& = & \frac{\sigma_0}{2} (V_\mathrm{AB} + V_\mathrm{BA}) + \sigma_xV_\mathrm{AA} + \frac{\sigma_z}{2}(V_\mathrm{AB}-V_\mathrm{BA}).
\label{S1}
\end{eqnarray}
where \(S_x\), \(S_0\), and \(S_z\) represent the contribution to the interlayer potential from, respectively, AA stacking, the average of AB and BA stacking, and the difference of AB and BA stacking.

In Fig.~\ref{pot2} we show these potentials for a pure screw (left hand panel) and a mixed partial dislocation (right hand panel). For the pure screw partial dislocation we show the potentials obtained from values of the tight-binding constant $B_\perp$ that sets the decay length of the interlayer interaction. As can be seen, the interlayer potentials are quite sensitive to the form of the tight-binding interaction. For a sufficiently fast decay of the interlayer interaction (the full lines in panel (1) of Fig.~\ref{pot2}) the only hopping allowed is between two atoms directly on top of each other. As a consequence during the transition region from AB to BA stacking the interlayer interaction becomes substantially weakened at the the core of the dislocation, as can be seen from the pronounced dip in the stacking average potential $S_0$. The AA component of the interlayer interaction, $S_x$, remains zero throughout the transition. For a slower decay of the interaction, on the other hand, one finds that the stacking average potential remains nearly unchanged with the AA potential finite throughout the transition region from AB to BA and showing a maxima at the dislocation core. Interestingly, although the \(S_x\)- and \(S_0\)-potentials depend strongly on the range of the interlayer hopping, the stacking difference potential \(S_z\) is quite insensitive to it.

One can also note that for the pure screw partial dislocations shown in panel (1) of Fig.~\ref{pot2} the interlayer potentials are all real. For screw dislocations aligned along the high symmetry armchair directions of the lattice this is always so, and using the general result for the interlayer potential given by Eq.~\ref{interlayer_coupling} it is easy to show this is related to the mirror symmetry that the lattice possesses about the dislocation core.

For the lattice geometry $\v a_1=(0,1)a$, $\v a_2 = (\frac{1}{2},\frac{\sqrt{3}}{2})a$; there are three armchair directions in the $\hat{\v y}$ direction and $\hat{\v y}$ rotated by $\pm 2 \pi/3$. The reciprocal lattice vectors thus have three lines of mirror symmetry: the $k_x$ line and this line rotated by $\pm 2 \pi/3$. Consider the mirror symmetry through the $k_x$ line, then every $\v G_j$ vector has a partner with $k_y$ component of opposite sign. For a $\Delta \v u(x) = \v f(x) \hat{\v y}$ form of the deformation field, i.e. one acting in the armchair $\hat{\v y}$ direction then terms in the lattice sum always come in complex conjugate pairs: the dot product in the phase $e^{-\mathrm{i}\Delta \v{u}(\v{r})\cdot\v{G}_j}$ ensures the mirror symmetry related $k_y$ components survive while the $k_x$ component vanishes identically. As the coefficients of these phases are identical for elements of each star, this proves the reality of the interlayer potentials for this case, and similarly for the other two armchair directions.

\subsection{Zero mode}

For real potentials, we can decompose the Hamiltonian into the tensor product

\begin{equation}
H=\tau_0 \otimes \sigma_x p_x +  \tau_3 \otimes \sigma_y p_y +\tau_3\otimes \sigma_0 \Delta+ \tau_x \otimes S(x).
\end{equation}
If we now perform a unitary rotation in layer space

\begin{equation}
U_1= R_y(\pi/2)\otimes \sigma_0=\frac{1}{\sqrt{2}}\begin{pmatrix}
1&-1\\
1&1
\end{pmatrix}
\otimes
\sigma_0
\end{equation}
that sends \(\tau_z \rightarrow -\tau_x\), \(\tau_x \rightarrow\tau_z\) and \(\tau_y\rightarrow \tau_y\) we find

\begin{equation}
H_1=U_1^\dagger H U_1 = \begin{pmatrix}
\sigma_x p_x + S(x) &-\sigma_y p_y - \Delta\\
-\sigma_y p_y -\Delta  & \sigma_x p_x -S(x)
\end{pmatrix}
\end{equation}
We next apply to $H_1$ the gauge transformation \(U_2\)

\begin{equation}
U_2=\begin{pmatrix}
\Phi_-&0\\
0&\Phi_+
\end{pmatrix}
\otimes\sigma_0
\end{equation}
where \(\Phi\) is given by

\begin{equation}
\Phi_\pm (x) = e^{\pm\mathrm{i}\int_0^x S_x(x')\mathrm{d}x'}
\end{equation}
The derivative in the \(x\)-momentum will bring down a factor of \(\pm S_x\) and together with the \(\sigma_x\) matrix from the momentum this will cancel the corresponding term in \(S(x)\).
After these manipulations the Hamiltonian takes the form

\begin{equation}
H_2=\begin{pmatrix}
\sigma_x p_x + \sigma_0 S_0+\sigma_z S_z & -(\sigma_y p_y +\Delta) \Phi_-^2\\
-(\sigma_y p_y  + \Delta)\Phi_+^2 & \sigma_x p_x - \sigma_0 S_0 -\sigma_z S_z
\end{pmatrix}
\end{equation}

\vspace{0.5cm}
We now simplify the Hamiltonian by setting \(p_y\) and the interlayer bias potential \(\Delta\) to zero, decoupling the two diagonal blocks. The average stacking has a value close to $1/2$ across the dislocation, representing therefore simply an energy shift $\pm 1/2$ between the two decoupled blocks. Neglecting this constant shift in energy we find

\begin{equation}
H_2'=
\begin{pmatrix}
\sigma_x p_x + \sigma_z S_z & 0\\
0 & \sigma_x p_x - \sigma_z S_z
\end{pmatrix}
\end{equation}

If we now apply another unitary rotation, this time in sublattice space,

\begin{equation}
U_3=\tau_0
\otimes\frac{1}{\sqrt{2}}
\begin{pmatrix}
1&- \mathrm{i}\\
-\mathrm{i}&1
\end{pmatrix}
\end{equation}
we send \(\sigma_z\rightarrow\sigma_y\) finding

\begin{equation}
H_3=\begin{pmatrix}
h_+ & 0\\
0 & h_-
\end{pmatrix}
:=\begin{pmatrix}
\sigma_x p_x +\sigma_y S_z & 0\\
0 & \sigma_x p_x -\sigma_y S_z
\end{pmatrix}
\end{equation}
The zero energy eigenvector of this Hamiltonian is 

\begin{equation}
\ket{\phi} = \begin{pmatrix}
\ket{\phi_+}\\
\ket{\phi_-}
\end{pmatrix}
\end{equation}
where \(\phi_\pm\) are given by

\begin{align}
\ket{\phi_+} &= e^{-\int_0^x S_z(x') \mathrm{d}x'} \ket{-}\\
\ket{\phi_-} &= e^{-\int_0^x S_z(x')\mathrm{d}x'} \ket{+}
\label{zeroMode}
\end{align}
\(\ket{\pm}\) is the eigenvector of \(\sigma_z\) with eigenvalue \(\pm\). It is easy to check that they satisfy the equation \(h_\pm \phi_\pm =0\). The only condition \(S_z\) must fulfil is that for the wave function to be normalizable the stacking difference potential must asymptotically change sign across the partial dislocation line.

\subsection{Full Spectrum}

To determine the full spectrum we employ the trick of squaring the Hamiltonian. 
After squaring \(H_3\) we find

\begin{equation}
H_3^2=\begin{pmatrix}
h_+^2 & 0 \\
0 & h_-^2
\end{pmatrix}
=
\begin{pmatrix}
p_x^2+S_z^2+\partial_x S_z & 0 & 0 & 0\\
0 & p_x^2+S_z^2-\partial_x S_z & 0 & 0\\
0 & 0 & p_x^2 + S_z^2 - \partial_x S_z\\
0 & 0 & 0 & p_x^2+S_z^2+\partial_x S_z
\end{pmatrix}
\end{equation}
 
We now choose a hyperbolic functional form \(\frac{1}{2}\tanh\left(\frac{x}{2l}\right)\)  for \(S_z(x)\). The corresponding Schr\"odinger equation for \(h_+^2\) gives us 

\begin{equation}
	\left[-\partial_x^2 + \frac{1}{4}\tanh^2\left(\frac{x}{2l}\right) + \sigma_z\frac{1}{4l}\sech^2\left(\frac{x}{2l}\right)\right]\Phi=E^2\Phi
\end{equation}
where \(l\) is the length of the dislocation (measured in units of \(l_\perp\)). After rescaling \(x:=\frac{x}{2l}\) and rearranging terms we find

\begin{equation}
	\left[-\partial_x^2-l(l-\sigma_z)\sech^2x\right]\Phi =(4E^2l^2-l^2)\Phi.
\end{equation}
We recognise the second term on the left hand side as the P\"oschl-Teller potential

\begin{equation}
V(x) = - \lambda (\lambda + 1) \sech^2(x)
\end{equation}
for \(\lambda = l\) and \(\lambda = l-1\).
The eigenvalues of the P\"ochl-Teller Hamiltonian are \(\epsilon =-m^2= 4E^2l^2-l^2\) and this immediately gives us the spectrum of the unsquared Hamiltonian

\begin{equation}
E=\pm\frac{\sqrt{l^2-m^2}}{2l}
\end{equation}
for integer \(m\) and \(l\).

The solutions to the P\"oschl-Teller potential are just the associated Legendre polynomials of the hyperbolic tangent function.

\begin{equation}
\Phi=P^m_l(\tanh x)=\frac{(-1)^m}{2^l l!}(1-\tanh^2 x)^{\nicefrac{m}{2}}\frac{\mathrm{d}^{m+l}}{\mathrm{d}^{m+l}(\tanh x)}(\tanh^2x-1)^l
\end{equation}
and we can now use these solutions of the squared Hamiltonian to construct the solutions of the original unsquared Hamiltonian. It turns out that the off-diagonal entries of \(h_+\) and \(h_-\)

\begin{equation}
h_+=\begin{pmatrix}
0 & -\mathrm{i}\partial_x - \frac{\mathrm{i}}{2}\tanh\left(\frac{x}{2l}\right)\\
-\mathrm{i}\partial_x + \frac{\mathrm{i}}{2}\tanh\left(\frac{x}{2l}\right) & 0
\end{pmatrix}:=\begin{pmatrix}
0&\mathrm i a\\
\mathrm i a^\dagger&0
\end{pmatrix}
\end{equation}
act as lowering and raising operators for the associated Legendre polynomials 

\begin{align}
aP_l^m&=-\frac{l+m}{2l}P_{l-1}^m (\tanh(x/2l))\\
a^\dagger P_{l-1}^m&=\frac{l-m}{2l}P_l^m(\tanh(x/2l))
\label{railow}
\end{align}
and so we make the Ansatz

\begin{equation}
\Psi =\begin{pmatrix}
c_+ P_{l-1}^m\\
c_- P_{l}^m
\end{pmatrix}
\end{equation}
for the eigenvectors of \(h_+\). The coefficients can easily be obtained from the eigenvalue equation giving us the eigenvectors corresponding to energies \(E\):

\begin{equation}
\Psi_l^m(\tanh(x/2l))=\frac1N\begin{pmatrix}
\sqrt{l+m}P_{l-1}^m\\
\pm\mathrm{i}\sqrt{l-m}P_{l}^m
\end{pmatrix}
\end{equation}
This expression, however, holds only for finite energies. If we set \(m=l\), i.e. \(E=0\), the second component vanishes due to the square root factor while in the first component we get \(P_{l-1}^l\), which is also zero. This results from the fact that in  the derivation of the spectrum we relied, in using Eq.~\ref{railow}, on the fact that \(l\neq  m\). We thus now instead use the fact that \(a \Psi_l^l = 0\) to build the ground state

\begin{equation}
\Psi_0 = \frac{1}{N}\begin{pmatrix}
0\\
P_l^l
\end{pmatrix}
\label{ptzero}
\end{equation}
We can confirm this by inserting \( \frac{1}{2}\tan\left(\frac{x}{2l}\right)\) into the general expression for the zero mode (Eq.~\ref{zeroMode}) yielding

\begin{equation}
\Psi_0 = \frac{1}{N}e^{-\frac{1}{2}\int_0^x \tanh(x'/2l) \mathrm{d}x'}\ket{-} =\frac{1}{N} \frac{1}{\cosh^l(x/2l)}\ket{-} =\frac1N \begin{pmatrix}
0\\
P_l^l
\end{pmatrix}
\end{equation}
We have thus found that at a dislocation between AB and BA stacking there exists, in addition to the valley Hall edge states, a series of of mass inversion states localised on the edges, the Volkov-Pankratov states. Their energies lie inside the valence and conduction bands while their number depends on the width of the dislocation. For each \( l\) there are \(2\lfloor l\rfloor +1\) bound states.

\begin{figure}[t!]
\centering
\includegraphics[width=0.45\textwidth]{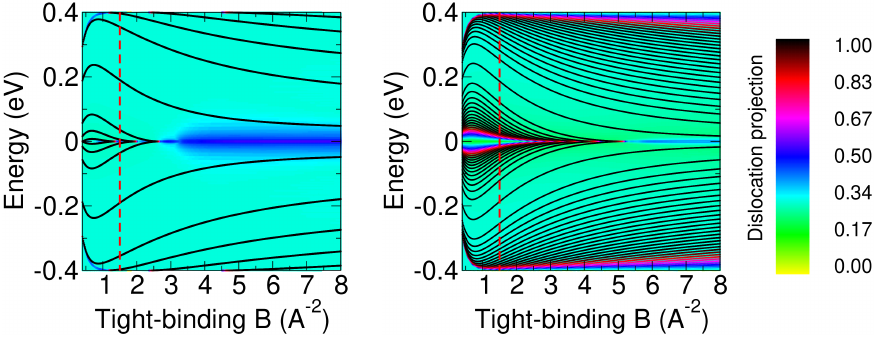}
\caption{Evolution of Volkov-Pankratov states at the $\Gamma$ point as a function of range of the interlayer potential, shown on the left hand side for a dislocation width of $l=50a$ and on the right hand side for an exaggerated stacking transition $l = 350a$. Close to $B=1.5$ (indicated by the vertical dashed line) one finds a spectrum showing (i) a mid-gap state at 0.2~eV and (ii) a symmetric energy spectrum of states either side of this mid-gap state, a situation close to that of the dislocation spectrum obtained from the P\"oschl-Teller potential.}
\label{fig1}
\end{figure}

\section{Dependence of Volkov-Pankratov spectrum on interlayer interaction}

We consider here the dependence on the Volkov-Pankratov (VP) spectrum on the interlayer interaction. By increasing the parameter $B_\perp$ we sweep through from a soft slowly decaying interlayer interaction to a "hard" fast decay. In Fig.~\ref{fig1} we show the VP spectrum at $p_y=0$ for a pure screw partial dislocation with $l=50a$, left hand panel, and $l=350a$ right-hand panel; VP states are seen for all values of $B_\perp$, with the P\"oschl-Teller spectrum described in Sec.~\ref{model} found only close to $B_\perp\sim 1.5$\si{\angstrom}$^{-2}$.

\subsection*{Acknowledgments}

All authors thank DFG for funding:
BM for funding through SFB 953 “Synthetic Carbon Allotropes” (project number 182849149), F.W. for the Graduate School GRK 2423 (project number 377472739, Sharma for TRR227 (ID 328545488, project A04), and Shallcross for funding through grant number SH 498/4-1.

%\bibliographystyle{unsrt}
%\bibliography{literature}


\end{document}